\documentclass{aa}
\usepackage{graphicx}
\begin{document}
   \title{Metallicity of mono- and multiperiodic $\beta$ Cephei stars}

   \author{J. Daszy\'nska-Daszkiewicz, E. Niemczura
          }

   \offprints{J. Daszy\'nska-Daszkiewicz}

   \institute{Astronomical Institute of Wroc{\l}aw University,
              ul. Kopernika 11, 51-622 Wroc{\l}aw, Poland\\
              \email{daszynska@astro.uni.wroc.pl}
             }
   \date{Received ...; accepted ...}
   \abstract{ Analyzing IUE ultraviolet spectra of $\beta$ Cep pulsating
            stars we have noticed that multiperiodic
            variables have a larger mean metal abundance in the
            photosphere, [m/H], than monoperiodic ones.
            We apply statistical tests to verify this dichotomy.
            We obtain that, with a large probability, the multiperiodic
            $\beta$ Cep stars have greater values of [m/H].
            This result is consistent with the linear
            non-adiabatic theory of pulsation of early B-type stars.
    }

   \titlerunning{Metallicity of mono- and multiperiodic $\beta$ Cephei stars}
   \authorrunning{J.Daszy\'nska-Daszkiewicz, E. Niemczura}
   \maketitle

\section{Introduction}

The $\beta$ Cephei pulsators are a well-known and well studied
group of early-type pulsating stars. Oscillations in these
stars are strictly connected with the metal abundance $Z$, as
they are driven by the classical $\kappa$-mechanism operating in the
layer of the metal opacity bump $(T\approx 2\cdot 10^5~{\rm K})$
caused by the lines of Fe-group elements (Moskalik \& Dziembowski
1992, Dziembowski \& Pamyatnykh 1993, Kiriakidis et al. 1992 and
Gautschy \& Saio 1993). The size of the instability domain of
$\beta$ Cep stars is very sensitive to the heavy element
abundance $Z$, and is smaller for lower values of $Z$
(Dziembowski \& Pamyatnykh 1993, Pamyatnykh 1999). Therefore, the
information about the metal abundance in $\beta$ Cep stars is of
special importance.

In previous works (Daszy\'nska 2001, Daszy\'nska et al. 2003)
we derived the metallicity parameter in the atmosphere, [m/H],
and the mean stellar  parameters for all $\beta$ Cephei stars
monitored by the {\it International Ultraviolet Explorer}.
We showed that the metallicity values obtained are independent
of effective temperature and surface gravity. Additionally,
they are not correlated with the stellar rotation and any
of the pulsational parameters, like a dominant period or
amplitudes of the light and the radial velocity variations.

In the scope of her PhD thesis, Daszy\'nska (2001) suggested that, in
general, the multiperiodic $\beta$ Cep stars have higher values of
[m/H] than the monoperiodic ones. We mentioned this result in
Daszy\'nska et al. (2003), and we confirmed it in Niemczura \&
Daszy\'nska-Daszkiewicz 2004 (Paper I), where we applied a more
accurate method of determination of stellar parameters. Using the
results of Paper I, we verify here this hypothesis by means of
statistical tests.

The paper is composed as follows. In Sect. 2 we remind derived
values of [m/H]. Sect. 3 contains the test of statistical
hypotheses about equality of two means. In Sect. 4 we discuss the
dichotomy of $\beta$ Cep variables with regard to the metallicity.
Conclusions are given in Sect. 5.

\section{Metallicity of $\beta$ Cephei variables from UV spectra}

In previous work (Paper I) we obtained the values of metallicity
in the atmosphere for 31 field $\beta$ Cep stars and 16 ones
belonging to three open clusters: NGC 3293, NGC 4755 and NGC 6231.
Our approach consisted of fitting theoretical fluxes of Kurucz
(1996) to low resolution IUE spectra, using the method based on
the least-squares optimalization algorithm (Bevington 1969).
Errors were estimated by the method of {\it bootstrap resampling}.
For more details about the method we refer the reader to Niemczura (2003).

We analyzed IUE spectra from two calibrations: NEWSIPS and INES.
The estimated values of the sample mean, standard deviation and
standard error of the mean, for [m/H] derived from these two date
sets, are\\[0.3cm]
NEWSIPS:
$${\overline {\rm [m/H]}}= -0.134, ~s_{\rm [m/H]}=0.201, ~s_{\overline {\rm
[m/H]}}=0.029$$
INES:
$${\overline {\rm [m/H]}}=-0.125, ~s_{\rm [m/H]}=0.203, ~s_{\overline {\rm [m/H]}}=0.030.$$
As we have shown in Paper I, these two populations follow a normal
distribution. Applying the two sample t-Test and the one-way ANOVA
test, we found that the two means are not significantly different
up to the significance level $\alpha=0.83$. Also, these two
data sets give the same dichotomy of $\beta$ Cep stars with regard
to the value of metallicity: higher [m/H] for multiperiodic
pulsators and lower [m/H] for monoperiodic pulsators. Therefore,
from now on, we will focus only on the NEWSIPS results.

For the purpose of this work, we remind the obtained values of [m/H].
In Table 1 we put the field $\beta$ Cep stars, while in Table 2 -
$\beta$ Cep stars belonging to the three open clusters. The
following columns contain the HD number, the star name, the value of
the dominant period and the value of [m/H]. The information about
mono- or multiperiodicity at the present knowledge is given in the
last column (Y=yes, N=no).

\begin{table}
\begin{flushleft}
\caption{The values of [m/H] for the field $\beta$ Cep stars.}
\begin{tabular}{rccrcc}
\hline
 HD~~ & Name & P [d] & [m/H]~~~ &  Multi. \\
\hline
  886  & $\gamma$ Peg & 0.15175 & $-0.04\pm0.08$  & N \\
\hline
16582  & $\delta$ Cet & 0.16114 & $-0.24\pm0.09$  & N \\
\hline
29248   & $\nu$ Eri   & 0.17351 &  $0.05\pm0.09$  & Y \\
\hline
44743  & $\beta$ CMa  & 0.25002 &  $0.04\pm0.10$  & Y \\
\hline
46328  & $\xi^1$ CMa  & 0.20958 & $-0.33\pm0.19$  & N \\
\hline
50707  &  15 CMa      & 0.18456 &  $0.05\pm0.10$  & Y \\
\hline
52918  &  19 Mon      & 0.1912  & $0.15\pm0.10$   & Y \\
\hline
56014  &  27 CMa      & 0.0918  &  $-0.47\pm0.18$   & N \\
\hline
111123 & $\beta$ Cru  & 0.19120 & $-0.14\pm0.13$   & Y \\
\hline
116658 & $\alpha$ Vir & 0.1738 &  $-0.12\pm0.08$   & N  \\
\hline
118716 & $\epsilon$ Cen & 0.16961 & $-0.14\pm0.10$ & Y \\
\hline
120307 & $\nu$ Cen   & 0.195 & $-0.26\pm0.10$  & Y  \\
\hline
122451 & $\beta$ Cen  & 0.157 & $-0.03\pm0.15$ &  Y \\
\hline
126341 & $\tau^1$ Lup & 0.17735 & $-0.39\pm0.16$  &  N \\
\hline
129056 & $\alpha$ Lup & 0.25985 &  $0.04\pm0.10$  &  Y \\
\hline
129557 &   BU Cir     & 0.12755 &  $-0.27\pm0.11$  &  Y \\
\hline
129929 &  V836 Cen    & 0.14313 &  $-0.05\pm0.10$  &  Y \\
\hline
132058 & $\beta$ Lup & 0.232 & $-0.35\pm0.11$  & Y  \\
\hline
136298 & $\delta$ Lup & 0.1982 & $-0.25\pm0.11$   &  N \\
\hline
144470 & $\omega^1$ Sco & 0.067 & $-0.45\pm0.15$ & N  \\
\hline
147165 & $\sigma$ Sco & 0.24684 &  $-0.20\pm0.20$  &  Y \\
\hline
157056 & $\theta$ Oph & 0.14053 & $-0.15\pm0.12$  &  N \\
\hline
158926 & $\lambda$ Sco & 0.21370 & $-0.21\pm0.08$ &  Y \\
\hline
160578 & $\kappa$ Sco & 0.19987 & $-0.19\pm0.10$  &  Y \\
\hline
163472 &  V2052 Oph   & 0.13989 & $-0.25\pm0.16$  &  N \\
\hline
199140 &  BW Vul     & 0.20104 &  $0.07\pm0.12$  &  N \\
\hline
203664 &   SY Equ   & 0.16587 &  $-0.01\pm0.21$ &  N \\
\hline
205021 & $\beta$ Cep & 0.19049 &  $-0.07\pm0.10$ &  Y \\
\hline
214993 & 12 Lac     & 0.19308 &  $-0.20\pm0.10$  &  Y \\
\hline
216916 & 16 Lac     & 0.16917 & $-0.13\pm0.13$  &  Y \\
\hline
       &  HN Aqr    & 0.15218 &  $0.21\pm0.09$ &   N  \\
\hline
\end{tabular}
\end{flushleft}
\end{table}
\begin{table}
\begin{flushleft}
\caption{The values of [m/H] for the $\beta$ Cep stars in
clusters.}
\begin{tabular}{rccrc}
\hline
 Name~~ & ID & P [d] & [m/H]~~~  & Multi \\
\hline
 {\bf NGC 3293}  & &  & &   \\
\hline
V380 Car & N3293-27 & 0.2273 & $0.06\pm0.10$  & Y \\
\hline
V400 Car & N3293-11 & 0.1458 & $-0.10\pm0.12$ & Y \\
\hline
V401 Car & N3293-10 & 0.1689 & $-0.14\pm0.12$ & Y \\
\hline
V404 Car & N3293-23 & 0.1621 &  $0.27\pm0.07$ & Y \\
\hline
V405 Car & N3293-14 & 0.1524 & $0.15\pm0.11$  & Y \\
\hline
V412 Car & N3293-65 & 0.1135 & $0.03\pm0.11$  & Y \\
\hline
 {\bf NGC 4755} &  & &  & \\
\hline
BW Cru &   N4755-F & 0.203 & $-0.31\pm0.15$  & Y \\
\hline
      &   N4755-G  & 0.156 &  $-0.56\pm0.20$ & N \\
\hline
      &   N4755-I  & 0.179 & $-0.29\pm0.10$  & N \\
\hline
      & N4755-I-13 & 0.232 & $-0.49\pm0.12$ & N \\
\hline
    & N4755-III-01 & 0.130 & $-0.50\pm0.18$  & N \\
\hline
 {\bf NGC 6231} &  & &  &  \\
\hline
V947 Sco & N6231-110 & 0.1079 & $-0.03\pm0.23$ & Y \\
\hline
V920 Sco & N6231-150 & 0.1012 & $-0.21\pm0.19$ & Y \\
\hline
V964 Sco & N6231-238 & 0.0878 & $0.05\pm0.15$ & Y \\
\hline
V946 Sco & N6231-261 & 0.0988 & $0.01\pm0.13$ & Y \\
\hline
         & N6231-282 & 0.1193 & $-0.42\pm0.16$ & Y \\
\hline
\end{tabular}
\end{flushleft}
\end{table}

Let us divide the sample of $\beta$ Cep stars in two groups:
monoperiodic $\beta$ Cep stars (a random variable $X$) and
multiperiodic ones (a random variable $Y$). These samples have
sizes $n=17$ and $m=30$, respectively. The sample parameters in
these two groups are the following
$${\bar x}=-0.252,~s_x=0.221,~s_{\bar x}=0.054,$$
$${\bar y}=-0.068,~s_y=0.156,~s_{\bar y}=0.029.$$
The same parameters obtained from Gaussian fitting are
$${\bar x}=-0.258,~s_x=0.223, ~s_{\bar x}=0.056,$$
$${\bar y}=-0.055,~s_y=0.188, ~s_{\bar y}=0.046.$$
Distributions of monoperiodic and multiperiodic stars as a
function of [m/H], together with Gaussian fits and normal curves,
are shown in Fig.1. In the left panel, the distribution of [m/H]
for monoperiodic $\beta$ Cep stars is shown, whereas in the right
panel the distribution of [m/H] for multiperiodic $\beta$ Cep
stars is shown.

   \begin{figure*}
   \centering
    \includegraphics[width=\textwidth,clip]{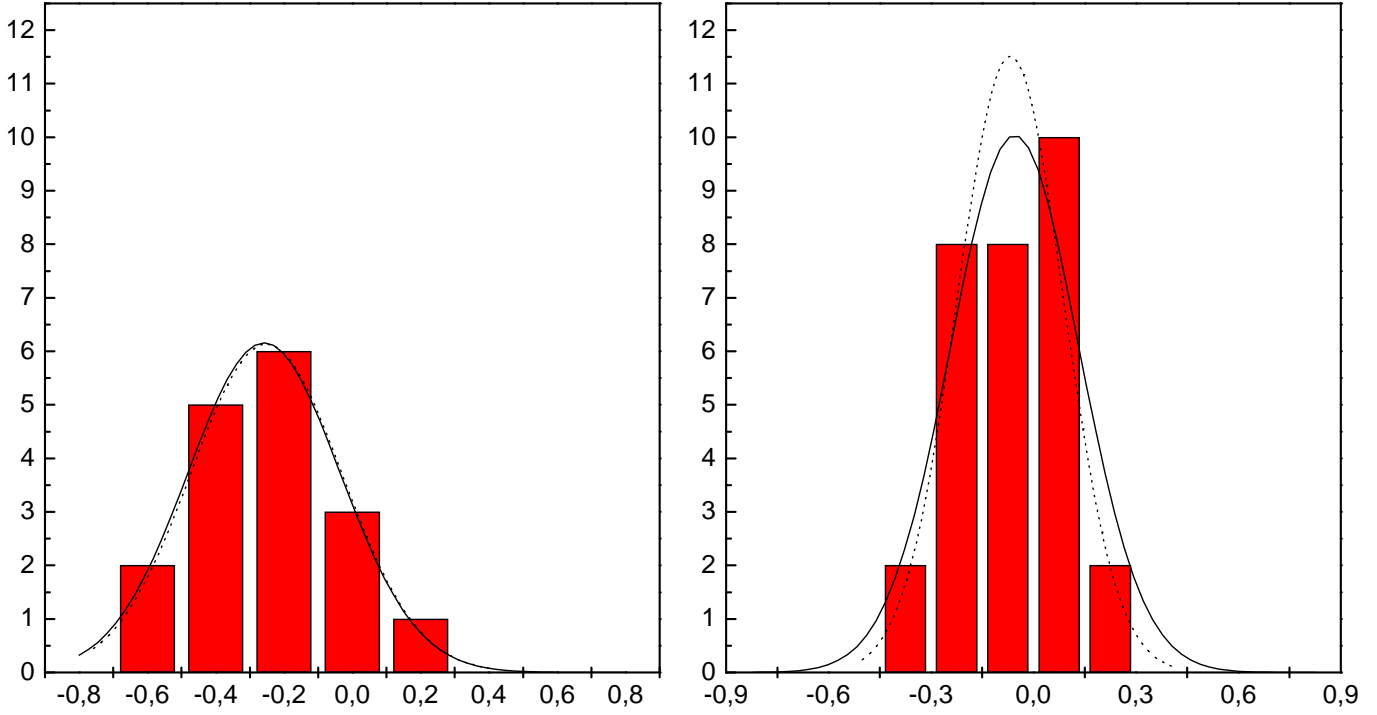}
           \caption{The distributions of [m/H] for monoperiodic $\beta$ Cep stars (left panel)
                  and for multiperiodic ones (right panel). The normal distribution (solid line)
                  and the Gaussian fit (dotted line) are also shown. }
        \label{aaaaaa}
         \end{figure*}

Comparing the sample parameters with those derived from Gaussian
fitting we can say that they are the same within the error limits.
Additionally we checked the normality of these distributions by
using the Shapiro-Wilk test. Thus, we may assume that $X$ and $Y$
have independent normal distributions $N(\mu_X,\sigma^2_X)$ and
$N(\mu_Y,\sigma^2_Y)$, respectively, where $\mu$ and $\sigma^2$
are the mean and the variance, respectively.

\section{Statistical tests}

We are interested in comparing the means of the metallicity
of the monoperiodic and multiperiodic $\beta$ Cep stars,
called the random variables $X$ and $Y$, respectively.
Because the sample sizes are small, we have to do that in two steps.
Firstly, we have to test whether the variances of these two variables
are equal, and then, if we accept this hypothesis, we will test the
equality of the two means.

\subsection{Test for the variances}

The null hypothesis $H_0:~\sigma^2_X=\sigma^2_Y$ will be tested against the
alternative hypothesis $H_1:~\sigma^2_X\neq\sigma^2_Y$. For this purpose
we use the statistic
$$F=\frac{(n-1)S_X^2/(n-1)\sigma^2_X} {(m-1)S_Y^2/(m-1)\sigma^2_Y}
= \frac{S_X^2}{S_Y^2}$$
which has an $F$-Snedecore distribution with $r_1=n-1$ and
$r_2=m-1$ degrees of freedom. Thus, $F_{\alpha}(n-1,m-1)$ is
needed to obtain the $\alpha$ significance level. $S_X^2$ and
$S_Y^2$ are assumed to be unbiased estimates of the corresponding
variances, $\sigma^2_X$ and $\sigma^2_Y$, of two independent
normal distributions with unknown means. To test the hypothesis
$H_0:~\sigma_X^2=\sigma_Y^2$ against a two-sided hypothesis
$H_1:~\sigma_X^2\neq\sigma_Y^2$, we assume the critical region
given by
$$\frac{S_X^2}{S_Y^2}\ge F_{\alpha/2}(n-1,m-1)$$
and
$$\frac{S_Y^2}{S_X^2}\ge F_{\alpha/2}(m-1,n-1)$$
Taking $s_x^2$ and $s_y^2$ as the observed values of unbiased
estimates of the variances, we get $F=2.01$.
At the significance level $\alpha=0.05$, the critical region, based on
the random samples of size $n=17$ and $m=30$, is
$$s_x^2/s_y^2\ge F_{0.025}(16,29)~~~{\rm and}~~~s_y^2/s_x^2\ge F_{0.025}(29,16)$$
or
$$F_{0.025}(16,29)\le s_x^2/s_y^2 \le \frac1{F_{0.025}(29,16)}=F_{0.975}(16,29)$$
hence
$$2.30 \le s_x^2/s_y^2 \le 0.39$$
Because our value of $F=2.01$ does not belong to the critical
region $(0,0.39>\cup<2.30,\infty)$, there are no reasons to reject
the hypothesis $H_0:~\sigma^2_X=\sigma^2_Y$. In this case, the
p-value, i.e. the smallest level of significance, $\alpha$, at
which $H_0$ would be rejected, amounts to 0.83. Thus, with a very
high probability we may assume the equality of variances.

\subsection{Test for the means}

Given that the hypothesis of equal variances is accepted, we are
ready to test the hypothesis about the means. If random
samples of small sizes $n$ and $m$ are taken, the test can be
based on the statistic
$$T=\frac{{\bar X}-{\bar Y}}{S_P\sqrt{1/n+1/m}},$$
where
$$S_P=\sqrt{\frac{(n-1)S_X^2+(m-1)S_Y^2}{n+m-2}}.$$
The $T-$statistic has a Student's $t$ distribution with $r=n+m-2$
degrees of freedom. The critical regions for testing the null
hypothesis $H_0:~\mu_X=\mu_Y$ against the alternative ones are

\begin{enumerate}

\item[(a)] $|t|\ge t_{\alpha/2}(n+m-2)~~{\rm
if}~~H_1:~\mu_X\neq\mu_Y,$

\item[(b)] $t\le -t_{\alpha}(n+m-2)~~{\rm if}~~H_2:~\mu_X<\mu_Y,$

\item[(c)] $t\ge t_{\alpha}(n+m-2)~~{\rm if}~~H_3:~\mu_X>\mu_Y.$
\end{enumerate}

We remind that the random variable $X$ represents [m/H] for the
monoperiodic stars, and the random variable $Y$ represents [m/H]
for the multiperiodic ones. The case $(a)$ means that we test whether
the mean values of [m/H] for the monoperiodic $\beta$ Cep stars and
the multiperiodic ones are equal or not. In the case $(b)$ we
verify the hypothesis of equality of these means against the
hypothesis that the mean for monoperiodic variables is less
than for the multiperiodic ones. And the last one, $(c)$, is the
test whether the monoperiodic stars have a greater value of [m/H]
than the multiperiodic ones.

If ${\bar x},~{\bar y}$ and $s_x^2,~s_x^2$ represent the observed
unbiased estimates of the respective parameters $\mu_X,~\mu_Y$ and
$\sigma^2_X=\sigma^2_Y$ of two independent normal distributions,
then the empirical value of the test statistic is $t=-3.33$. At
$\alpha=0.05$ and $n+m-2=45$ we have the following critical
regions for these three alternative hypotheses

\begin{enumerate}

\item[(a)] $(-\infty,-2.014>\cup<2.014,\infty),~~{\rm
if}~~H_1:~\mu_X\neq\mu_Y$,

\item[(b)] $(-\infty,-1.680>~~{\rm if}~~H_2:~\mu_X<\mu_Y$,

\item[(c)] $<1.680,\infty)~~{\rm if}~~H_3:~\mu_X>\mu_Y.$
\end{enumerate}

The values of the observed $t$ belong to the first and the second
critical regions. Thus $H_0$ is clearly rejected at an
$\alpha=0.05$ significance level in favor of the alternative
hypothesis $H_1:~\mu_X\neq\mu_Y$, as well as in favor of the
hypothesis $H_2:~\mu_X<\mu_Y$. On the other hand we would not
reject the hypothesis $H_0$ if the alternative one is
$H_3:~\mu_X>\mu_Y$. In this case we have to accept that
$\mu_X=\mu_Y$, but it is impossible because $H_0$ has already
been rejected.

The $p-$value for the two-side hypothesis, $H_1:~\mu_X\neq\mu_Y$, is
0.002, and for the one-side hypothesis, $H_2:~\mu_X<\mu_Y$, amounts to
0.001. Thus, with a large probability, on the basis of these data
sets, we can conclude that the mean metallicity for multiperiodic
$\beta$ Cep stars is higher than that for monoperiodic ones.

\section{Discussion of the metallicity dichotomy}

The result from our statistical analysis is in agreement with the
linear non-adiabatic theory of pulsation of $\beta$ Cep stars. As
was shown by Dziembowski \& Pamyatnykh (1993), for greater metal
abundance we get more unstable modes. In Fig. 2 we plot the
evolution of oscillation frequencies in the Main Sequence phase
for four indicated masses and two values of the heavy element
abundance $Z$. We considered low degree modes with $\ell=0,1,2$.
In these computations we used the Warsaw-New Jersey stellar evolution
code and the non-adiabatic pulsation code of Dziembowski (1977).
We took OPAL opacities and did not include effects of rotation
and convective core overshooting. As we can see from Fig.2,
for the higher $Z$ value we get a wider range of unstable frequencies.

Guided by this property, we plot in Fig.3 two dependencies, taking
all studied multiperiodic $\beta$ Cep stars. In the top panel we
show the range of the observed frequencies as a function of [m/H],
whereas in the bottom panel, the range normalized by the sum of
the lowest and highest frequency as a function of [m/H]. In both
cases we did not find any correlation. The correlation
coefficients, ${\tilde\rho}$, calculated including errors as weights,
are equal to -0.31 and -0.45 for the frequency range
and the normalized frequency range, respectively.
   \begin{figure}
   \centering
\includegraphics[bb=70 156 388 743, width=88mm,clip]{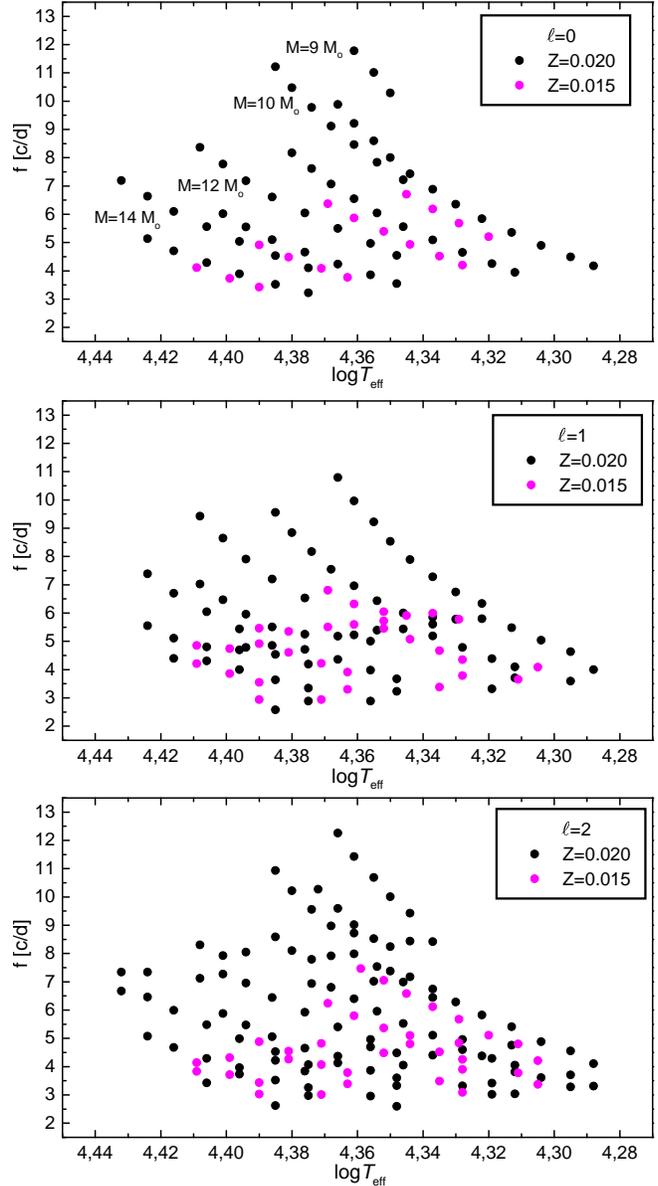}
          \caption{Frequencies of unstable modes as a function of effective temperature
                    in the MS sequence phase of evolution for four values of mass:
                    $M=9,10,12,14 M_{\sun}$ and two values of the heavy element
                    abundance: $Z=0.020$ and $0.015$. Three panels correspond to three
                    values of the spherical harmonic degree, $\ell$.}
        \label{aaaaaa}
         \end{figure}
   \begin{figure}
   \centering
    \includegraphics[width=88mm,clip]{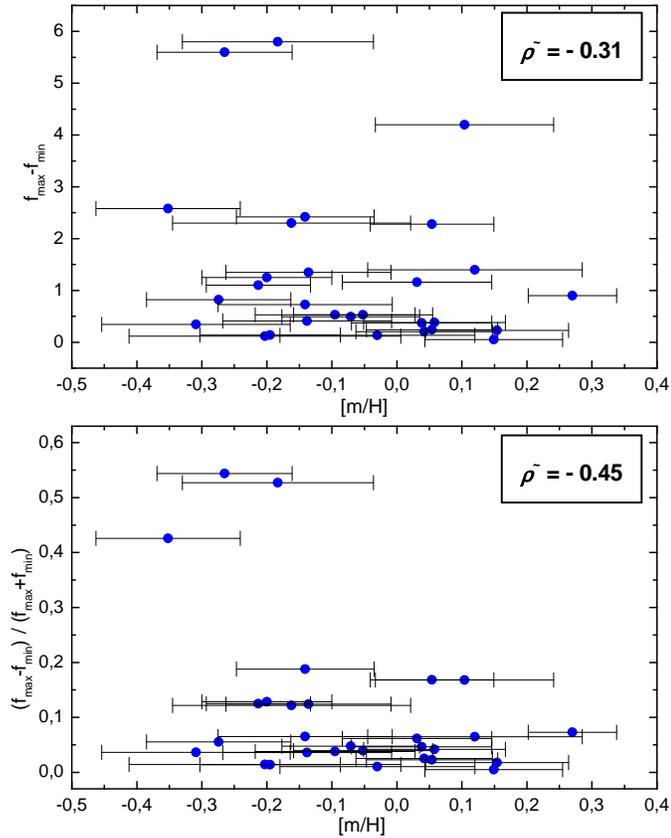}
     \caption{The range of observed frequencies in $\beta$ Cep stars as a function of [m/H].
              The bottom panel is for the frequency range normalized by the sum of the lowest
              and highest frequency. The values of weighted correlation coefficient, $\tilde\rho$,
              are given in the right-upper corners.}
        \label{aaaaaa}
         \end{figure}

This result is not surprising. $\beta$ Cep stars are pulsating
variables with small amplitudes both in photometry as well as in
spectroscopy, and we do not know how many modes we do observe
amongst the excited ones. The number of known frequencies
increases every year. This is mainly thanks to large multi-site
and multi-technique campaigns, organized already for many $\beta$
Cep variables. For example, the last campaign of $\nu$ Eri
(Aerts et al. 2004, Handler et al. 2004) gave an increase of the
frequency range from $~0.14$ up to $~2.28$. The data from the 12
\& 16 Lac campaign set up by Gerald Handler are already analyzed.
Great hopes are pinned also on forthcoming space mission, e.g.
the COROT project. Therefore the dependencies in Fig. 3 are just a
first trial, which should be constantly updated.

\section{Conclusions}

Using statistical tests, we have shown that with a very high
probability the multiperiodic $\beta$ Cephei stars have a higher
mean metallicity in their atmospheres than the monoperiodic ones.
The statistical  analysis was based on our last determinations of
the metallicity parameter [m/H] (Paper I). Getting a normal
distribution of [m/H] for the mono- and multiperiodic stars, we
applied the test about equality of variances of these two samples,
represented by X and Y random variables. The test based on the $F$
statistic allowed clearly to accept the null hypothesis
$H_0:~\sigma^2_Y=\sigma^2_Y$. Then we could test the hypothesis
about the mean values of metallicity for the mono- and
multiperiodic variables. We used the $T$ statistic to define the
critical regions for a small size of random samples. The null
hypothesis $H_0:~\mu_X=\mu_Y$ was rejected in favor of
$H_1:~\mu_X\neq\mu_Y$, as well as in favor of $H_2:~\mu_X<\mu_Y$.

The higher value of the mean metal abundance in the case of
multiperiodic $\beta$ Cep stars is understandable in the framework
of the linear non-adiabatic theory, which predicts more unstable
frequencies for higher metal abundance $Z$. Therefore, we tried
to find a dependence between the range of observed frequencies and
the metal abundance. We can not make a statement about any correlation
between these two parameters, because the value of the weighted correlation
coefficient $\tilde\rho$ amounts to -0.45. In fact, we should
not expect more, as a number of detected modes is much lower than
theory predicts. We have to remember also that the values of [m/H]
give information mainly about photospheric metal abundances.
Moreover, we do not take into account phenomena like
diffusion, element mixing etc.

To improve this analysis, a few things must be done. Firstly, we
need determinations of the [m/H] parameter for a larger sample of
$\beta$ Cep stars. Secondly, a high accuracy in the period search
analysis is needed to estimate better the range of the excited modes
in a given star. Here put our hope on next multi-site and multi
technique campaigns, as well as on space missions. Finally, the
analysis of the chemical composition of $\beta$ Cep variables from
high resolution spectra would be very helpful.

These results require further studies and must be treated with
caution. Firstly, they were obtained for small samples of stars,
and secondly, we do not know how many modes from the excited ones
have been already identified. Our aim was only to show some
properties which could be extracted from such determinations of
the metal abundance parameter [m/H] for $\beta$ Cep variables.
Having in mind that from a theoretical point of view all $\beta$ Cep
stars should be multiperiodic ones, the metallicity dichotomy we
obtained is rather a qualitative confirmation of the dependence of
the pulsation instability in these stars on the heavy element
abundance.

\acknowledgements{We thank Wojtek Dziembowski and
Alosha Pamyatnykh for the use of the Warsaw-New Jersey
evolutionary code and the non-adiabatic pulsation code.}


\begin{thebibliography}{}

  \bibitem[2004]{aerts2004} Aerts, C., De Cat, P.; Handler, G. et al., 2004, MNRAS, 347, 463

  \bibitem[1969]{bevington} Bevington, P.R. 1969, in Data Reduction and Error Analysis for the Physical Science,
   New York: McGraw-Hill

  \bibitem[2001]{daszynska01} Daszy\'nska, J. 2001, PhD Thesis, Wroc{\l}aw University, Poland

  \bibitem[2001]{daszynska03} Daszy\'nska, J., Niemczura E. \& Cugier, H. 2003, Adv. Space Res. 31, 387

  \bibitem[1977]{dziem93} Dziembowski, W. A., 1977, Acta Astron., 27, 203

  \bibitem[1993]{dziem93} Dziembowski, W. A. \& Pamyatnykh, A. A. 1993, MNRAS, 262, 204

  \bibitem[1993]{gautschy93} Gautschy, A. \& Saio, H. 1993, MNRAS, 262, 213

  \bibitem[2004]{handler2004} Handler G., Shobbrook R.R., Jerzykiewicz M., et al., 2004, MNRAS, 347, 454

  \bibitem[1992]{kiriakid92} Kiriakidis M., El Eid M.F., Glatzel W. 1993, MNRAS, 255, 1P

  \bibitem[1996]{kurucz96} Kurucz, R.L. 1996, CD-ROM No. 13 and 19

  \bibitem[1992]{moskal1992} Moskalik P., Dziembowski W.A. 1992, A\&A, 256, L5

  \bibitem[2003]{niemczura03} Niemczura E., 2003, A\&A 404,689

  \bibitem[2003]{niemczura03} Niemczura E., Daszy\'nska-Daszkiewicz J., 2004, Paper I, A\&A

  \bibitem[1993]{pamyat99} Pamyatnykh, A. A. 1999, Acta Astron. 49, 119

\end{thebibliography}
\end{document}